\documentclass{article}
\pdfoutput=1

\usepackage{PRIMEarxiv}

\usepackage[utf8]{inputenc} 
\usepackage[T1]{fontenc}    
\usepackage{hyperref}       
\usepackage{url}            
\usepackage{booktabs}       
\usepackage{amsfonts}       
\usepackage{nicefrac}       
\usepackage{microtype}      
\usepackage{lipsum}
\usepackage{fancyhdr}       
\usepackage{graphicx}       
\graphicspath{{media/}}     
\usepackage{authblk}

\usepackage{bbm, dsfont}
\title{Mapping Inequalities in Activity-based Carbon Footprints of Urban Dwellers using Fine-grained Human Trajectory Data
}
\author{}
\author[1*]{Akhil Anil Rajput \thanks{akhil.rajput@tamu.edu}}
\author[1]{Yuqin Jiang\thanks{yuqinjiang@tamu.edu}}  
\author[2]{Sanjay Nayak \thanks{sanjaynayak@tamu.edu}}  
\author[1]{Ali Mostafavi\thanks{amostafavi@civil.tamu.edu}}
\affil[1]{Zachry Department of Civil and Environmental Engineering, Texas A\&M University, College Station, TX 77840, USA}
\affil[2]{Department of Computer Science and Engineering, Texas A\&M University, College Station, TX 77843, USA}
\affil[*]{Corresponding author}

\begin{document}
\maketitle




\section*{Abstract}

Effective climate mitigation strategies in cities rely on understanding and mapping urban carbon footprints. One significant source of carbon is a product of lifestyle choices and travel behaviors of urban residents. Although previous research addressed consumption- and home-related footprints, activity-based footprints of urban dwellers have garnered less attention. This study relies on deidentified human trajectory data from 5 million devices to examine the activity-based carbon footprint in Harris County, Texas. Our analysis of the heterogeneity of footprints based on places visited and distance traveled reveals  significant inequality: 10\% of users account for 88\% of visitation-based footprints and 71\% of distance-traveled footprints. We also identify the influence of income on activity-based carbon footprint gap of users related to their travel behavior and lifestyle choices, with high-income users having larger footprints due to lifestyle choices, while low- to medium-income users' footprints are due to limited access. Our findings underscore the need for urban design adjustments to reduce carbon-intensive behaviors and to improve facility distribution. Our conclusions highlight the importance of addressing urban design parameters that shape carbon-intensive lifestyle choices and facility distribution, decisions which have implications for developing interventions to reduce carbon footprints caused by human activities.

\section{Introduction}

\begin{figure*}[!ht]
\centering
\includegraphics[width=0.8\textwidth]{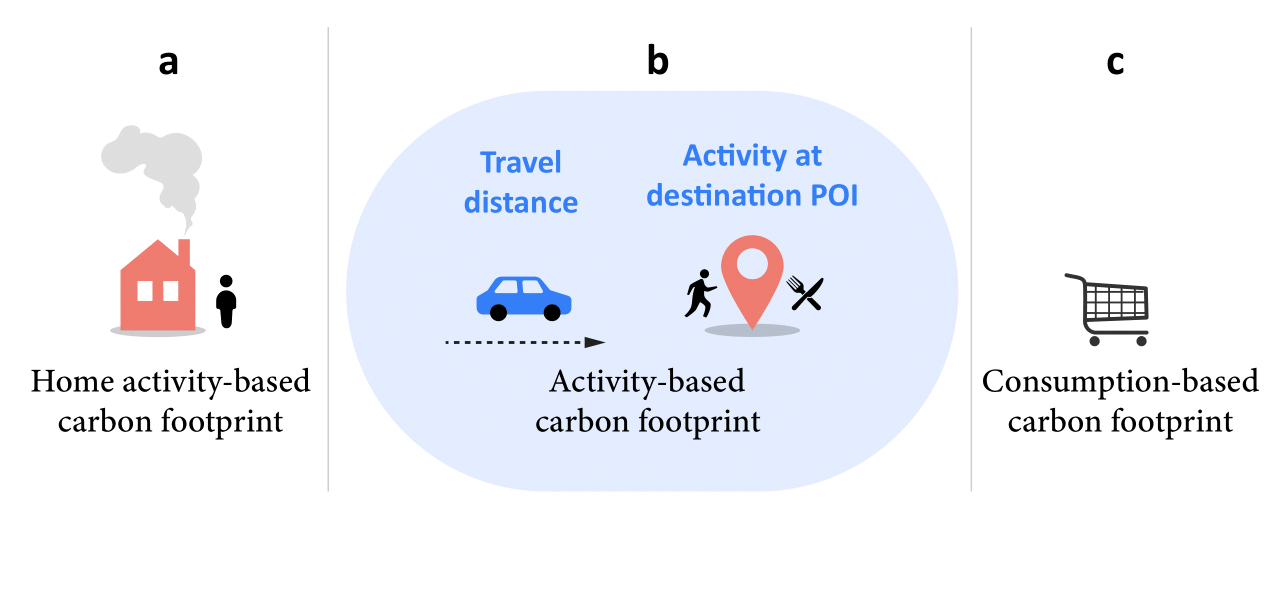}
\caption{Illustration of the components of residents' life-activity carbon footprint: (a) home-activity-based footprint, is related to in-house activities such as energy consumption (b) activity-based footprint, which accounts for the emissions generated during travel between home and destination points of interest; visitation-based footprint, representing emissions resulting from activities during visitation to destination points of interest; and (c) consumption-based carbon footprint, which encompasses the emissions associated with the consumption of goods and services during home activities. This study focuses on activity-based carbon footprints of individuals.}
\label{fig:introduction}
\end{figure*}

Cities function as hubs of economic growth and social transformation; however, rapid urbanization  challenges the attainment of environmental sustainability and climate mitigation \cite{ayres1991greenhouse, heidari2016review}. Normal life activities contributing to a city's carbon footprint comprise three components (Fig. \ref{fig:introduction}): (1) home activity-based footprint; (2) activity-based footprint, and (3) consumption-based footprint. Home-activity-based carbon footprint captures consumption of any form of energy used within the home \cite{goldstein2020carbon}. Activity-based carbon footprint refers to an individual's interaction with the built environment through travel outside their homes \cite{salo2021drivers, smetschka2019time}. Consumption-based carbon footprint refers to all products consumed or used in the course of daily life. Carbon emissions are generated during the production and transportation of these goods \cite{ivanova2016environmental, ivanova2017mapping, zhang2016novel}. Among these three components of residents' life activity carbon footprint, the literature has paid far greater attention to home-activity-based and consumption-based carbon footprints; thus, our understanding of activity-based carbon footprints is rather limited. Activity-based carbon footprint is driven primarily by residents' lifestyle and travel patterns. The literature shows that patterns of human lifestyle mobility in cities are influenced by urban forms and structures \cite{ccolak2016understanding, rajput2022anatomy, esmalian2022characterizing}. Thus urban design and development, such as facility distribution, can influence activity-based carbon footprints. Accordingly, understanding ramifications of patterns of carbon footprints associated with individuals' activities and mobility in urban areas can inform decision makers and urban planners in the development of targeted strategies and interventions to reduce urban residents' carbon footprint and to promote more effective mitigation strategies. 

Studies examining activity-based carbon footprints use survey-based data to learn of activities individuals engage in; however, time-based assessments from survey data fail to fully capture patterns of human visitation to points of interest and the associated travel distance, both of which influence the extent of carbon footprint. This gap can be addressed through the use of observational fine-grained human trajectory data. Accordingly, this study seeks to address four research questions: (1) To what extent do activity-based carbon footprints of individuals vary across different residents?; (2) To what extent does an activity-based carbon footprint gap exist in cities?; (3) What proportion of individuals account for the majority of activity-based carbon footprint in cities?; and (4) To what extent do activity-based carbon footprint profiles of individuals vary based on income? To answer these research questions, we investigate the heterogeneity in activity-based carbon footprint (including both visitation-based footprint and distance traveled by users) in Harris County (Houston metropolitan area), Texas. Our analysis is based on high-resolution user-level waypoint data collected by INRIX, a location-based data provider; building polygons from Microsoft; points of interest location and attribute data from  SafeGraph; and US Census data. By examining the patterns of carbon emissions associated with visitation and distance traveled, our goal is to shed light on the factors contributing to the heterogeneity in users' carbon footprints and their implications for urban design and development, as well as for climate mitigation strategies. It is important to note that although the total carbon footprint of an individual depends on an activity-based footprint derived from traveling and visitation to POIs, home activities, and consumption, our study focuses only on evaluating activity-based footprint, for which little knowledge and prior work exists.

Our results reveal intriguing patterns in the distribution of activity-based carbon footprints based on visitation and distance traveled, indicating substantial heterogeneity (inequality) among users. We observed that a small percentage of users account for the majority of contributions to activity-based carbon footprint. This heterogeneity may be influenced by factors such as city structure, lifestyle patterns influenced by income, distribution of facilities, and accessibility to POIs and facilities. The results show that, unlike consumption- and home-activity-based footprints, gaps in which the primary contributors are high-income individuals \cite{feng2021household, miehe2016regional, moran2018carbon}, low- and medium-income groups are among the highest contributors to high activity-based carbon footprint. The high activity-based footprint of high-income individuals, however, is attributable mainly to their lifestyle choices (type of POIs visited), while the high activity-based footprint of low-income individuals is due to the need for traveling longer distances to POIs to obtain necessities. These findings show that the significant carbon footprint gap among individuals in cities can inform integrated urban design strategies in which urban development plans and projects could target interventions for reducing carbon footprint gap to achieve climate mitigation goals in cities. 

The remainder of this paper is organized as follows: Section 2 provides information on relevant literature related to carbon footprint evaluation in cities, and Section 3 describes the data and methods used in our analysis, including the datasets, data preprocessing, and the analytical approach. Section 4 presents the results, focusing on the heterogeneity in distance traveled and POI visitation footprint, as well as the disproportionate impact of user groups on activity-based carbon footprint and distance traveled. Section 5 discusses the implications of our findings for understanding the drivers of carbon emissions and informing policy and planning efforts to promote sustainable urban development. 

\section{Related Work}

\subsection{Carbon Footprints of Individuals and Households}

Households and individuals play a significant role in global carbon emissions \cite{PATEL2022104087, ameli2015determinants, long2022carbon, CHENG2022104047}. Research has shown that households contribute to over half of national greenhouse gas emissions in various countries \cite{long2018policy, wiedenhofer2017unequal}. To understand and mitigate household-level carbon footprints, numerous studies have investigated carbon footprint patterns at the household or individual levels. Generally, carbon footprint at these levels can be categorized into direct and indirect emissions. Direct carbon footprint refer to direct energy consumption, such as heating, cooling, and electricity usage by appliances. Indirect emissions refer to greenhouse gases emitted by other entities producing goods and services households and individuals consume \cite{chen2022direct, epa2020, long2017exploring}. For instance, when an individual purchases a product, the indirect carbon footprint comprises the emissions generated during the production, transportation, storage, and sale of that product.

Previous studies have assessed the carbon footprint of individuals or households based on their expenditure records, taking into account both direct and indirect emissions that result from the consumption of goods and services by individuals or households. Expenditure-related studies relying primarily on survey data have been conducted in multiple countries, including China \cite{li2019impact, mi2020economic, wang2020urban, zhang2020intertemporal}, Japan \cite{koide2019carbon, long2017exploring}, Belgium \cite{Christis2019zr}, Germany \cite{Gill2018cv}, Norway \cite{isaksen2017carbon, steen2016carbon}, and the United States \cite{feng2021household, pottier2022expenditure, weber2008quantifying}. Expenditure-related studies focus primarily on consumption-related carbon footprint of households and individuals but do not capture activity-based carbon footprints.Multi-regional input-output (MRIO) models and related databases were also used extensively to understand consumption activities. MRIO data records economic flows across multiple regions or nations. The construction of MRIO tables allow researchers to track carbon emissions generated concomitant with economic flows \cite{fenner2018carbon, hasegawa2015carbon, lenzen2013building, leontief1986input, minx2009input, XING2022103977}.  

Another stream of research focuses on activity-based carbon footprints based on the time-use perspective. Time-based studies examine time as a unified unit, enabling the comparison of carbon footprint by different individual-level activities within the same time unit \cite{druckman2012time, jiang2022estimating, smetschka2019time}. Activity-based carbon footprint studies based on time-use perspective estimate carbon emissions levels for activities during the same time unit (usually an hour or fifteen minutes). With a common temporal duration, activities can be compared for their respective carbon footprint. Time-use perspective studies enable a better understanding of activity-based carbon footprints due to individual-level lifestyles \cite{jalas2015energy, druckman2019time}. This method has been implemented in studies in Japan \cite{jiang2022estimating, koide2019carbon}, China \cite{yu2020causal, yu2019time}, Finland \cite{jalas2015energy}, France \cite{de2017energy}, Austria \cite{smetschka2019time}, Britain \cite{druckman2012time}, and the United States \cite{sekar2018changes}. These studies found that activities such as eating out, personal care, and commuting, had high carbon footprints \cite{druckman2012time, druckman2019time, smetschka2019time, yu2020causal}. 

While the time-use studies have demonstrated the importance of capturing and analyzing activity-based carbon footprints, a significant limitation across time-use studies is the nature of their data collection methods. Most of these studies rely heavily on survey data, wherein participants self-report their time usage. Gathering time usage information through surveys is a labor-intensive process, and the self-reported data may not be accurate and representative, leading to an incomplete measurement of activity types and thus, inaccurate estimation of the activity-based footprint. To address these limitations, this study uses a human mobility dataset based on cell phone location service. Using this more comprehensive record of individuals' movement trajectories, the type and frequency of POIs visited, as well as distance traveled by each user can be accurately measured to compute the extent of activity-based carbon footprint (the total of visitation-based footprint and total distance traveled). Additionally, the human trajectory dataset provides data at a finer temporal resolution, enabling a deeper understanding of the characteristics of people's activity patterns and associated carbon footprints. 

\subsection{Carbon Footprint Gap}

Carbon footprint gap refers to the differences in the extent of carbon footprint of individuals and households. Various demographic and socioeconomic factors have been found to correlate with household carbon footprint, including income, gender, age, household composition, car ownership, employment \cite{long2022carbon, minx2013carbon, HUANG2022104236}. Generally, household income exhibits a strong positive correlation with carbon emissions, as higher-income households tend to have a higher demand for consumption. This relationship has been observed in Japan \cite{long2022carbon}, China \cite{liu2022challenges, wiedenhofer2017unequal}, Norway \cite{isaksen2017carbon}, and the United States \cite{feng2021household, jones2014spatial}. For the consumption-based footprint, the existing studies show a significant carbon footprint gap. The term carbon elites have been used to refer to higher-income households and individuals whose consumption of goods account for a greater portion of carbon footprint; however, regarding activity-based footprint, limited insights exist regarding the extent of carbon footprint gap and whether a similar carbon footprint elitism exists. 

Urban design can also affect how people interact with their surroundings, impacting their carbon footprint; however, urbanization presents mixed effects on carbon emissions \cite{ivanova2017mapping, wang2021impacts, zhang2017does}. On one hand, dense metropolitan areas can use less energy by encouraging people to use facilities like public transportation \cite{brown2009geography, ercan2016investigating, Gill2018cv, DORR2022104052}. The availability of a wider range of commodities and their ease of accessibility, on the other hand, might result in a rise in consumption, ultimately resulting in a rise in carbon emissions \cite{Gill2018cv, zhang2016novel}. The examination of activity-based carbon footprint of individuals can inform about differences among carbon footprints of individuals and possible associations with urban design characteristics. 

\section{Data and Methods}

This section provides an overview of the data used, the preprocessing approach for cleaning and merging the datasets, and the methods adopted for evaluating user carbon impact. In this study, we used human trajectory data from Harris County (Houston metropolitan area), Texas. We chose Harris county as a region of study for three reasons: first, Harris County is the most populous county in Texas and the third most populous in the United States \cite{nasser_2021}. Harris County also has a diverse population with a range of lifestyles and travel patterns. Second, Harris County is home to a large number and diverse type of points of interest (POIs), such as businesses, institutions, healthcare, and recreational facilities, which are relevant for measuring activity-based carbon footprint. Third, Harris County has a significant carbon footprint due to its high levels of transportation-related emissions \cite{ClimateActionPlan}, making it an important area for study in terms of carbon reduction and sustainability. 

Due to the volume of data required, privacy concerns, the complexity of analysis, and computational cost, we limited the study to one metropolitan city. Nevertheless, the methodology can be applied to any spatial area at any resolution. To evaluate user lifestyle and travel patterns, we collected anonymized waypoint travel data from INRIX \cite{inrix}. We also collected building polygons from Microsoft \cite{microsoftfootprintadmin_2020}, SafeGraph points of interest location and attribute data \cite{safegraphpoi}, and census data to link the travel stop locations to POIs or to a spatial unit, such as census tract. The integration of these diverse data sources allows delineation of types of POIs visited and distance traveled from each individual user's movement trajectory for a more holistic understanding of the relationship between user lifestyle, travel patterns, and activity-based carbon footprint.

\subsection{Datasets}

This study leverages a variety of unique datasets that facilitate the capture of human lifestyle and travel patterns to delineate activity-based carbon footprints. The primary dataset contains high-resolution user-level waypoint data, obtained from INRIX, a private location intelligence company that provides anonymized location-based data and analytics, ensuring the privacy of individuals. This dataset stands out for two reasons: first, it provides accurate coordinates for the trajectory for each trip, with INRIX collecting vehicle coordinates every few seconds, resulting in a high spatiotemporal resolution for a more granular analysis of travel patterns and their corresponding carbon footprints. Second, INRIX data was collected for the entire Houston metropolitan area (Harris County) over a three-month period (February through April 2016) at all times of the day, resulting in a dataset of more than 26 million trip records. 

To associate the stop locations from the waypoint data to POIs, we used Microsoft building footprint data. This dataset consists of building boundaries in a shapefile format, generated from Bing imagery using open-source toolkits and machine-learning techniques. While this dataset accurately identifies the building footprint or boundary, it does not identify the type of building or service it provides. To address this limitation, we incorporated the SafeGraph POI dataset, which offers precise POI coordinates for most brands and categorizes them by type of business. In this study, we focused on major categories to facilitate easier generalizability.

To categorize stop points corresponding to high-, medium-, or low-income neighborhoods, we collected US Census data from census.gov at the census-tract resolution. This data includes census-tract boundaries and the corresponding median household income for each area, which serves as a proxy for the income characteristics of that particular region. We used census data from the year 2020, assuming that income and major demographics remain relatively stable over time. This approach enables a more accurate understanding of the relationship between lifestyle, travel patterns, and carbon footprint across different income groups, ultimately informing about carbon footprint gap among various income groups.

\subsection{Data Preprocessing} 

\begin{figure*}[!ht]
\centering
\includegraphics[width=0.95\textwidth]{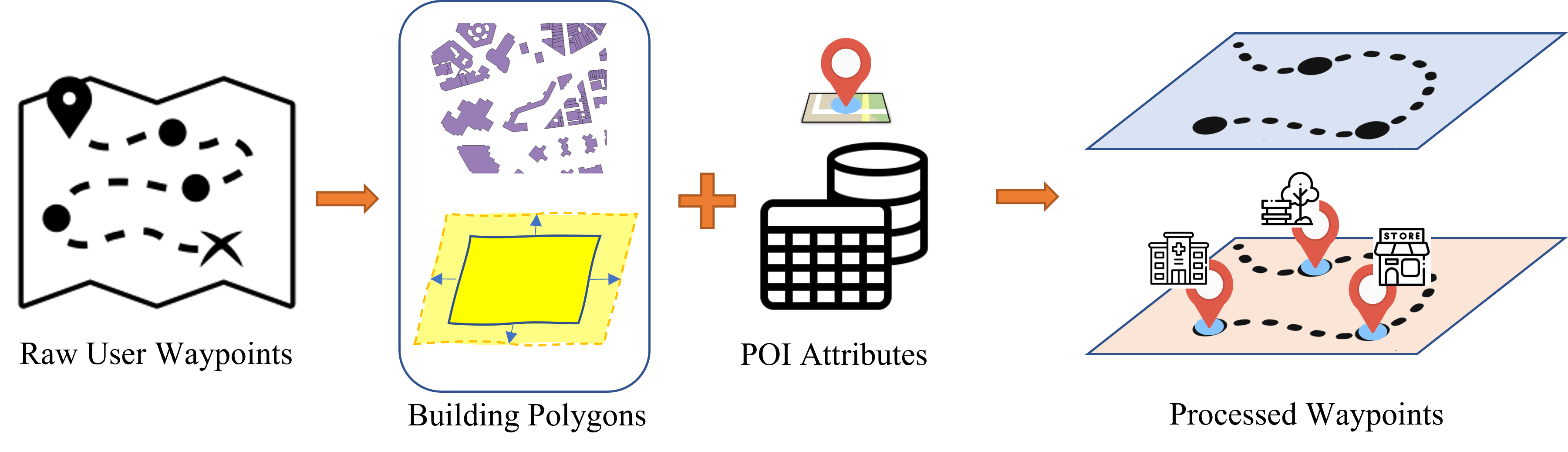}
\caption{Overview of the data preprocessing step. Raw user waypoints are merged and linked to building polygons and points of interest.}
\label{fig:pre_processing}
\end{figure*}

The data preprocessing stage was a crucial step in our study, involving several stages, including data cleaning, merging, and aggregation. We first focused on cleaning the high-resolution user-level waypoint data obtained from INRIX, which contained more than 26 million trip records for 6.5 million devices in Harris County spanning three months. The data cleaning process entailed the removal of trips that both originated and concluded outside our designated study region (Harris County) resulting in a dataset spanning 20 million trips across 5.16 million devices. Figure \ref{fig:pre_processing} illustrates the steps followed during preprocessing.

We first combined  SafeGraph POIs and building footprint data if a POI coordinate was located within a building footprint. This resulted in an enriched building footprint dataset containing additional attributes for POIs, such as brand name, POI type, category, subcategory, and other attributes along with the building boundary information. For the purpose of this study, we used only the category information for POIs, which classified POIs into high-, medium-, and low- footprint activities. This classification will be discussed in further detail in the next section. We then filtered out all building polygons that were not linked to any  SafeGraph POIs. These excluded building polygons likely represent residential buildings that are not relevant to this study.

The next step involved the linking of all the stop points from our waypoint data to building polygons by considering a 10-meter buffer around each building polygon. A buffer of 10 meters along the building footprint boundary was employed to account for stop points related to a POI that, due to GPS error, were not located within a building footprint. As the distance between the road system and POIs typically exceeds 10 meters, this method prevents the incorrect classification of stop points to POIs. In cases where the specific POI could not be determined for a stop point, such as brands inside a mall or a building with multiple stores, a random POI was selected to designate that stop point. This approach does not introduce bias, as most co-located POIs tend to belong to similar categories and thus have comparable visitation-based carbon footprints. We also linked stop points to census tract polygons if the stop points were situated within them, thereby associating stop points with different income categories based on the median household income in the census tract. 

Given the vast dataset comprising 26 million trip records, a total of 65,000 POIs, and 750 plus census-tract polygons that required linking through spatial operations such as buffering, the preprocessing stage was executed on a powerful 64-core and 500-GB RAM system to accommodate the high computational requirements for processing the data. One novel aspect of this study lies not only in the data preprocessing itself but in the optimization and parallelization of the code, which significantly reduced the computational time.

\subsection{POI Classification} 

As previously mentioned, individuals' carbon footprints are driven primarily by their activities, which in this study are categorized based on user visits to POIs and distance traveled. To evaluate users' carbon footprint according to their visitation activity, we first classified POIs into high, medium, and low carbon-footprint activities. These categorizations were primarily based on a framework developed by \cite{smetschka2019time}. We also referred to studies from \cite{yu2019time, druckman2012time, jiang2022estimating} that conducted similar categorizations. For instance, we designated outdoor activities, such as visiting a park, as low-carbon intensive since the activity itself does not significantly contribute to the carbon footprint. Conversely, activities such as dining at a restaurant were considered high-carbon footprint activities.

Following the data preprocessing stage, which involved linking each stop point in a trip to a POI if the stop point was in close proximity, we obtained approximately 160 categories of POIs. Based on their closest functional resemblance, these POIs were manually reclassified into 15 categories, as shown in Table \ref{tab:poi_carbon_footprint}. If a POI did not correspond to any of these categories, it was assigned an "others" classification, and the POI carbon footprint was labeled manually. After categorizing the POIs into these 17 groups, we used the total footprint values captured in studies \cite{smetschka2019time, yu2019time, druckman2012time, jiang2022estimating} to classify the POIs into low, medium, and high carbon footprints. To aggregate the footprint impact of each trip, we assigned numerical values of 1, 3, and 5 to low, medium, and high carbon footprint categories, respectively. The literature indicated that the footprint of activities in the high category was at least twice that of medium and low footprint activities. This approach ensures that a visit to a high-footprint POI does not equate to visits to a low- and medium-footprint POI. Further details will be discussed in the methods section.

\begin{table}
\centering
\begin{tabular}{|c|l|l|}
\hline
\textbf{S.No.} & \textbf{POI Category} & \textbf{Carbon Footprint} \\ \hline
1 & business              & low                       \\ \hline
2 & construction          & high                      \\ \hline
3 & eating                & high                      \\ \hline
4 & education             & low                       \\ \hline
5 & entertainment         & high                      \\ \hline
6 & fitness               & low                       \\ \hline
7 & gardening             & mid                       \\ \hline
8 & home-related          & mid                       \\ \hline
9 & hotel                 & high                      \\ \hline
10 & infrastructure        & high                      \\ \hline
11 & manufacturing         & high                      \\ \hline
12 & medical               & high                      \\ \hline
13 & outdoor               & low                       \\ \hline
14 & repairs               & mid                       \\ \hline
15 & self-care             & mid                       \\ \hline
16 & shopping              & low                       \\ \hline
17 & transportation        & high                      \\ \hline
\end{tabular}
\caption{POI Categories and their corresponding carbon footprint levels. POIs that did not fall into any of these 17 categories were labeled manually.}
\label{tab:poi_carbon_footprint}
\end{table}

\subsection{Methods} 

\begin{figure*}[!ht]
\centering
\includegraphics[width=0.95\textwidth]{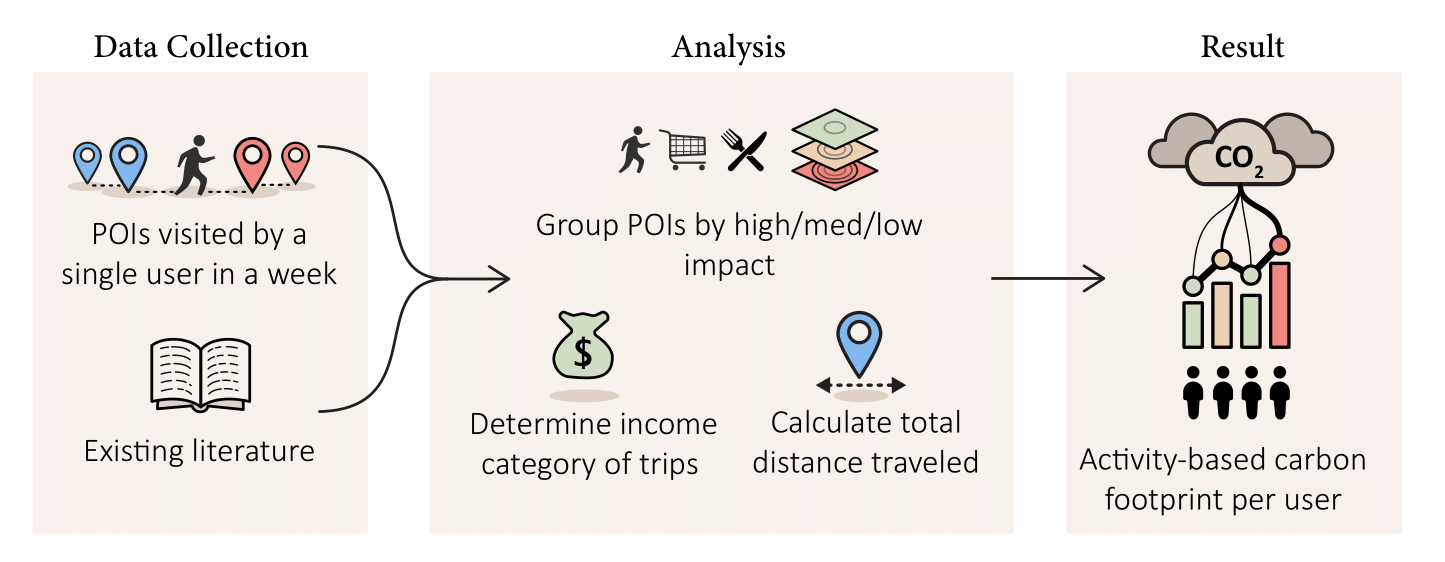}
\caption{Conceptual illustration of methods employed in the study. We first collected fine-grained waypoint data for users in Harris County and added location-based attributes to stop points, such as type of POI, activity-based footprint, average household income category, and distance traveled. Then we evaluated the spatiodemographic linkages between carbon footprint and other attributes.}
\label{fig:methods}
\end{figure*}

The preprocessing of the above-described data resulted in a refined dataset in which the stop points of every trip were linked to a POI, if in close proximity, and linked to the corresponding census tract to determine income attributes. With this refined dataset, we first evaluated the distance traveled in every trip made by a user. Next, we assessed the types of POIs visited and stop locations. By examining the types of POI visitations and frequency of visits, we gained insights into the visitation-based carbon footprint of each user and the income of the locality where these POIs visits or stop locations are situated. Figure \ref{fig:methods} depicts an overview of the steps for specifying visitation-based and distance-traveled carbon footprint of each user in this study.

In the next step, we aggregated all metrics at the weekly level for each user, meaning that for every user, a weeks' worth of trips were aggregated, summing up all the POIs visited and the total distance traveled. The weekly aggregation of POI visitations and distance traveled provide a typical activity-based footprint of users since the variations of POI visitations and trips across weeks are insignificant. Specific methods of aggregation will be discussed in more detail later in this section. This weekly aggregation approach was chosen to better capture the lifestyles of people in a city, as weekdays and weekends exhibit different mobility patterns and lifestyle activities \cite{yuan2012extracting, liu2009understanding, lenormand2014cross, ma2022characterizing}. By aggregating on a weekly basis, we encapsulated all the variability associated with the weekday–weekend effect, allowing for a more comprehensive representation of the overall lifestyle and behavior of users. 

\subsubsection{Visitation-based user carbon footprint}

From the preprocessed data, we have visitation records for each user's trip for the months of February through April. For each trip, we classified the POIs visited into high-, medium-, and low-footprint categories based on their categories, as discussed in the previous section. We then evaluated the effective carbon footprint for each user trip using the Equation 1:

\begin{equation}
CF_{v} = \sum_{i=1}^{n} w_{i} \times f_{i}
\end{equation}

where, $CF_{v}$ is the visitation-based carbon footprint for a trip made by a user, $w_{i}$ is the weight assigned to each POI category based on their carbon footprint, and $f_{i}$ is the frequency of POIs visited in each category during the trip.

The values of $w_{i}$ for low, medium, and high footprint categories are 1, 3, and 5, respectively.

For example, if a user visits two low-footprint POIs (such as a dog park and gym) and one high-footprint POI (such as a restaurant) during a trip, the total footprint will be seven (2x1 + 1x5 = 7). It should be noted that here that we account only for the footprint based on the types of POIs visited, so if a user does not visit any POI on that trip, then the visitation activity-based carbon footprint for that trip will be zero. We then aggregated these footprint values for each user for all trips at a weekly resolution. This means that a user's visitation-based carbon footprint will represent the weekly footprint that represents a user's lifestyle. Mathematically, it is represented by Equation 2:

\begin{equation}
CF_{w} = \sum_{j=1}^{m} CF_{v_j}
\end{equation}

where, $CF_{w}$ is the weekly visitation-based carbon footprint for a user, $CF_{v_j}$ is the visitation-based carbon footprint for the $j^{th}$ trip made by the user during the week, and $m$ is the total number of trips made by the user during the week.

\subsubsection{Distance-based carbon footprint}

Transport has direct and indirect carbon emissions which contribute to a user's carbon footprint. To account for this contribution in evaluation of activity-based footprints, we also calculated the distance traveled by each user in a trip and, subsequent total distance traveled by a user in a week. We first calculated the distance between consecutive stop locations using the Haversine formula, which considers the earth's curvature. We then summed up the distances between all consecutive stop locations to obtain the total distance traveled in a trip. Using this method, we then calculated the total distance traveled by each user in a week by aggregating the distances traveled in all trips. This approach allowed us to accurately estimate the distance-based carbon footprint of each user by considering their transportation activities in addition to their visitation-based carbon footprint. 

For a trip, the distance traveled by a user is calculated as the sum of the distances between consecutive stop locations using Equation 3:
\begin{equation}
DT_{trip} = \sum_{i=1}^{n-1} dist(loc_i, loc_{i+1})
\end{equation}

where $n$ is the total number of stop locations in a trip, $loc_i$ is the latitude and longitude of the $i^{th}$ stop location, and $dist(loc_i, loc_{i+1})$ is the distance between consecutive stop locations calculated using the Haversine formula.

For a week, the total distance travelled by a user is calculated by aggregating the distances traveled in all their trips using Equation 4:
\begin{equation}
DT_{week} = \sum_{j=1}^{m} DT_{trip_j}
\end{equation}

where $m$ is the total number of trips made by the user during a week, and $DT_{trip_j}$ is the distance traveled by the user in the $j^{th}$ trip.

By considering both the visitation activity-based carbon footprint and the distance-based footprint of users, we can obtain a comprehensive understanding of the users' overall activity-based carbon footprint for purposes of evaluating carbon footprint gap among users. 

\subsubsection{Activity-based carbon footprint gap}

To investigate the variation of activity-based carbon footprint across users and the extent of footprint gap, we utilized the median household income of the locations visited by a user as a proxy for the user's income level. We first determined the median household income for each location using publicly available data from the US Census Bureau. Then, for each user, we assigned an income category based on the median household income of the locations visited by the user. Users who visited locations with a median household income of less than 45,000\$ were categorized as low income, those with a median household income between 45,000–100,000\$ were categorized as median income, and those who visited locations with a median household income of more than 100,000\$ were categorized as high income. We did this by assigning an income category to each trip based on the category of the POIs of a particular income category visited with the highest frequency. 

To gain insights into the extent of the activity-based carbon footprint gap, we analyzed the distribution of users across different income categories and their corresponding carbon footprint and distance-traveled categories. This allowed us to identify any patterns or trends in the data, highlighting the variation of users' lifestyle choices and travel behavior across different income levels. We computed the proportion of low-, medium-, and high-income users in each of the low-, medium-, and high-visitation-based carbon footprint categories, as well as the proportion of users in each income group in the low-, medium-, and high-distance-based footprint categories. By comparing these proportions, we were able to discern whether users from different income groups have different activity-based carbon footprint profiles and to evaluate the extent of activity-based carbon footprint. We categorized the thresholds for high and low and the top 10 percentile and bottom 50 percentile of the respective visitation- and distance-based carbon footprint values. A linear split was not done as the values had an exponential distribution.

\section{Results}

\subsection{Heterogenity in distance traveled and POI visitation footprint}

The analysis results reveals intriguing patterns in the distribution of activity-based carbon footprints of urban dwellers based on visitation and distance traveled, indicating a substantial degree of heterogeneity and carbon footprint gap among users. 

\begin{figure*}[!ht]
\centering
\includegraphics[width=0.95\textwidth]{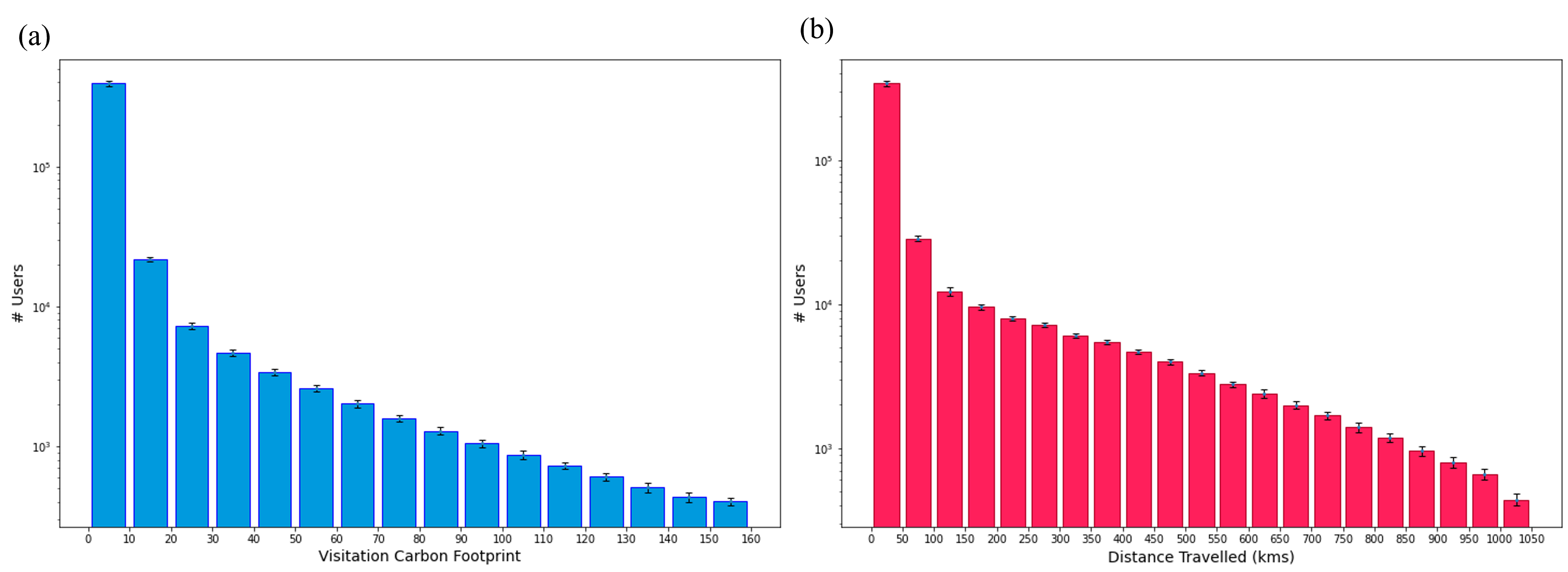}
\caption{Distribution of POI visitation-based and distance-based carbon footprint of users accounting for 99.9\% of activity. (a) Frequency plot of visitation-based carbon footprint versus number of users, displayed on a log scale. The plot shows the distribution of visitation-based carbon footprints for different frequency levels of users. The frequency of users with different carbon footprints follows a decaying trend that is more exponential in nature. (b) Frequency plot of distance-based carbon footprint versus number of users, displayed on a log scale. The plot shows the distribution of distance-based carbon footprints for different frequency levels of users. Similar to the visitation-based carbon footprint, the distribution of distance-based carbon footprints also follows a decaying trend that is more exponential rather than linear. The results indicate that both visitation-based and distance-based carbon footprints exhibit similar patterns of distribution with respect to the frequency of users.}
\label{fig:r1}
\end{figure*}

The results, as illustrated in Fig. \ref{fig:r1}, show that both visitation-based and distance-based carbon footprints exhibit similar patterns of distribution in terms of the frequency of users in different levels of carbon footprint. Fig. \ref{fig:r1} (a) presents the frequency plot of visitation-based carbon footprint versus the number of users on a log scale. The plot shows that the frequency of users decreases rapidly with higher levels of carbon footprint, suggesting that a small percentage of users has significantly higher contributions to visitation-based carbon footprint than the rest. The majority of users have a lower activity-based carbon footprint, but a few users are responsible for a disproportionately large share of visitation-based carbon footprint, highlighting the gap in the users' visitation-based carbon footprint. Similarly, Fig. \ref{fig:r1} (b) displays the frequency plot of distance-based carbon footprint versus the number of users on a log scale. The frequency of users also decreases exponentially with higher levels of distance-based carbon footprint. This result indicates that a small group of users travels longer distances and thus has a significantly higher distance-based carbon footprint. This finding further emphasizes the gap in users' activity-based carbon footprint, with a few users having the greatest activity-based carbon footprint.

These results suggest a high degree of heterogeneity and gap in users' activity-based carbon footprint (both visitation-based and distance-based), which has important implications for understanding and addressing the drivers of carbon emissions. By identifying the factors contributing to this heterogeneity, policymakers and urban planners can develop targeted interventions to promote more sustainable lifestyles and reduce the carbon footprint of urban residents. The heterogeneity in users' carbon footprints may be influenced by factors such as city structure, income, distribution of facilities, and accessibility to different modes of transportation.

\subsection{Activity-based carbon footprint gap across users}

\begin{figure*}[!ht]
\centering
\includegraphics[width=0.95\textwidth]{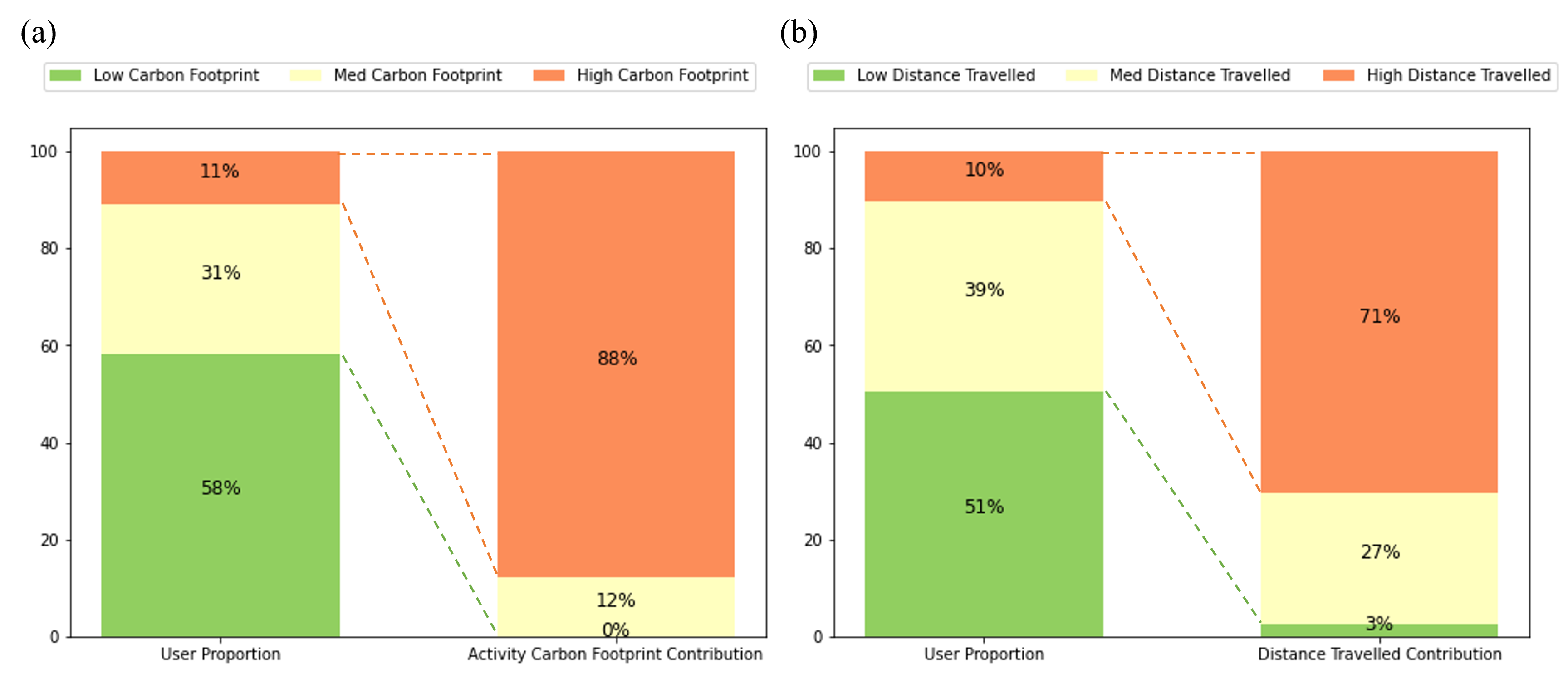}
\caption{Contribution of users to activity-based carbon footprint and distance traveled. (a) The stacked bar plot shows the contribution of users in high-, medium-, and low- footprint categories to the total activity-based carbon footprint. Among the users, 11\% are responsible for 88\% of the total carbon footprint, while 31\% contribute to 12\% of the total footprint. Interestingly, 58\% of the users have a carbon footprint of 0, indicating low or no impact on the carbon emissions from their activity. (b) The stacked bar plot depicts a similar trend with distance traveled. Here, 10\% of users account for 71\% of the total distance traveled, while 39\% contribute to 27\% of the total distance. Notably, 51\% of the users only have a minimal contribution of 3\% to the total distance traveled, suggesting a significant heterogeneity in the travel behavior and carbon emissions among the user population.}
\label{fig:r2}
\end{figure*}

The results from the stacked bar plot in Fig. \ref{fig:r2} highlight the contribution of users in different footprint categories to the total activity-based carbon footprint and distance traveled. In particular, as shown in Fig. \ref{fig:r2} (a), among the users, 11\% are responsible for a substantial 88\% of the total activity-based carbon footprint, indicating a disproportionately high impact on carbon emissions from their activity. Moreover, 31\% of users contribute to only 12\% of the total footprint, indicating a relatively lower impact. Interestingly, 58\% of the users have a carbon footprint of 0, suggesting a considerable portion of the population with minimal or no impact on carbon emissions from their activity. This could be due to visitations that do not correspond to travel to any POI, such as home to office and back. Also, one limitation of the dataset is that due to user anonymization .Device IDs may change, making it difficult to evaluate the weekly travel profile of an individual. This finding highlights the heterogeneity in the carbon footprint among users, with a small fraction of users contributing significantly to the total emissions, while a larger fraction has minimal impact. In terms of distance traveled by users in a week over all the trips, the stacked bar plot in Fig. \ref{fig:r2} (b) reveals a similar trend. 10\% of users account for 71\% of the total distance traveled, indicating a significant contribution to overall travel behavior and carbon emissions. On the other hand, 39\% of users contribute to only 27\% of the total distance traveled, suggesting a relatively lower impact. Notably, 51\% of the users only have a minimal contribution of 3\% to the total distance traveled, indicating a significant heterogeneity in travel behavior and associated carbon impact arising from the choice of travel across the user population. 

These findings suggest that there are distinct patterns of heterogeneity in both activity-based carbon footprint and distance traveled among users, which may be influenced by factors such as individual lifestyle choices, travel behavior, and activity patterns. Further analysis is needed to better understand the underlying factors driving these patterns of heterogeneity and their implications for carbon emissions in the context of user lifestyles in the city. A city's structure and layout can influence the distribution of facilities, services, and employment opportunities influencing the travel and lifestyle patterns of individuals. In cities with a more dispersed layout, users may need to travel long distances to access essential services or to reach their workplaces, leading to higher carbon emissions. To understand if the observed behavior is in part influenced by demographic attributes of the census tracts where the POIs exist, we also evaluated the dependence income on the distance traveled and activity-based carbon footprint in the next section.

\subsection{High footprint users: Choice or force?}

\begin{figure*}[!ht]
\centering
\includegraphics[width=0.95\textwidth]{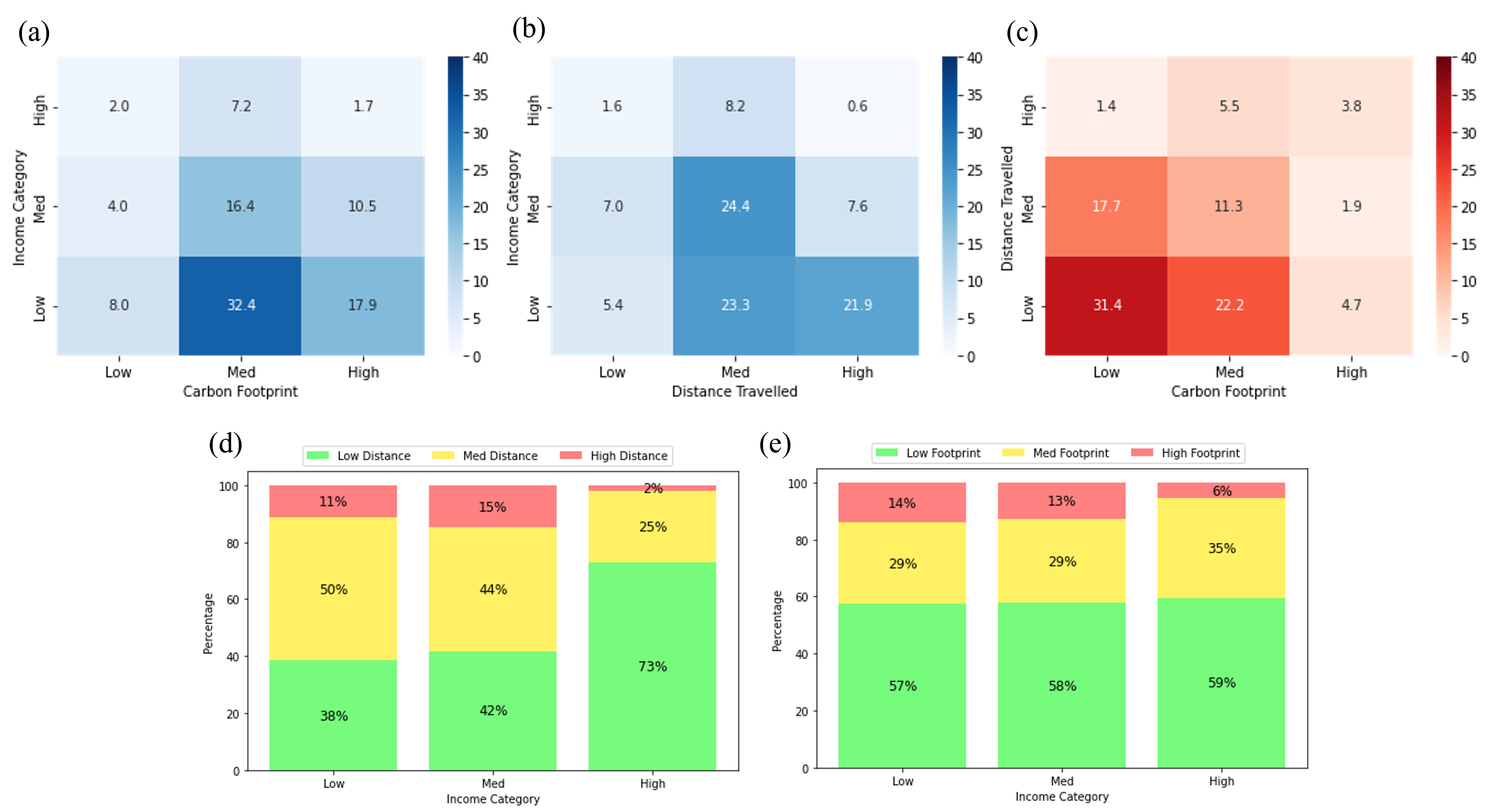}
\caption{Relationship between income, carbon footprint, and distance traveled by the users. (a) shows a heatmap representing the percentage of users in each of the nine categories, obtained by combining the three income categories with the three carbon footprint categories. The colors and values of the heatmap indicates the percentage of users falling in each of the nine categories. (b) and (c) show similar plots for income versus distance traveled and distance traveled versus carbon footprint, respectively. Plots (d) and (e) are stacked bar plots that illustrate the fraction of users in each income category that contribute to different distance-traveled categories and carbon-footprint categories, respectively.}
\label{fig:r3}
\end{figure*}

We evaluated the distribution of users across different income and visitation-based and carbon footprint and distance-traveled categories. We observed a significant difference in the visitation-based carbon footprint and distance traveled by users based on their income category. The heatmap plot in Fig. \ref{fig:r3} (a) shows the distribution of users across different income and visitation-based carbon footprint categories. It can be seen that a majority of high-income users fall into the low-carbon footprint category, while a higher percentage of low-income users fall into the high-carbon footprint category. This finding is intriguing, as one might expect that high-income users typically have more resources and opportunities to engage in activities with a higher carbon footprint, hence contribute more to carbon footprint. However, the results indicate that high-income users follow lifestyles that include more visits to low-footprint POIs. On the other hand, low-income users may visit POIs more frequently due to having fewer resources and a more constrained access. \cite{esmalian2022characterizing} showed that lower-income users visit grocery stores more frequently since they might not have resources to buy all their needs from the closest facility and in one visit. Greater frequency of POI visits and longer distance traveled translate into higher activity-based carbon footprints. Similarly, Fig. \ref{fig:r3} (b) shows the distribution of users across different income and distance-traveled categories. These results indicate that income significantly influences travel distances. Fig. \ref{fig:r3} (c) shows that a significant portion of users travel short or moderate distances while maintaining a low carbon footprint, suggesting better access for POIs contributing to low-footprint activities. On the other hand, some users have a high carbon footprint despite traveling short or moderate distances, which highlights the importance of interventions for reducing carbon-intensive lifestyle choices. 

Fig. \ref{fig:r3} (d) and (e) provide a clearer picture of the contribution of low, medium, and high-income categories to different distance traveled and visitation-based carbon footprint categories. It can be observed that a higher percentage of high-income users fell into the low-distance traveled category, while a higher percentage of low-income users fell into the high-distance traveled category. Interestingly only a fraction of users with high-income travel longer distances, which is more than five times less than low and medium-income category visits. This result could highlight the disparities in facility distribution and limited access in low-income neighborhoods requiring users to travel longer distances. Users in all income categories exhibit near-same proportion of low visitation-based footprint, indicating that a low visitation-based footprint is not strongly influenced by income category.
But users corresponding from higher-income areas contribute less than half to high visitation-based footprint compared to those of low and medium income. This could be partly due to the fact that high-income users may have greater flexibility in their lifestyle choices, such shopping locally, or engaging in low-carbon visitation activities, and also better access to facilities, which can contribute to reduced travel distances and a lower distance-based carbon footprint. On the other hand, low-income users may have limited choices and limited access, resulting in longer travel distances and a higher carbon footprint. Additionally, the uneven distribution of facilities in low-income neighborhoods may force users to travel longer distances to access essential services, amenities, and recreational activities, whereas high-income neighborhoods may have better access to such facilities, resulting in shorter travel distances. These observations suggest that urban design decisions related to facility distribution and access could influence the lifestyle of low-income populations in ways that force them to exhibit high activity-based carbon footprint behaviors. Hence, sustainable urban design strategies focusing on equitable facility distribution and improving access could also have positive effects on reducing the activity-based carbon footprint of lower-income residents \cite{wiedenhofer2018household, AYDIN2022103843, WANG2023104381, YANG2019783}. 

\section{Concluding Remarks}

Activity-based carbon footprint of urban dwellers is one of the least understood and studied components of carbon footprint analysis of urban dwellers. Addressing this important gap, in this study, we analyzed the users' activity-based carbon footprint using high-resolution waypoint data in the context of Harris County, Texas. Our findings provide valuable insights into understanding the user lifestyles that shape the extent of visitation-based and distance-based carbon footprints of individuals to evaluate carbon footprint gap among residents. Our results show disproportionately high activity-based carbon footprint from a small fraction of users, with 11\% of users responsible for 88\% of the total visitation-based carbon footprint, and 10\% of users accounting for 71\% of the distance-based carbon footprint. These results highlight a high degree of heterogeneity and a significant carbon footprint gap in both visitation-based and distance-based carbon footprint by users in Harris County, Texas. According to the results, a small percentage of urban residents have significantly higher contributions to activity-based carbon footprint, while a larger fraction of residents has the lowest footprint. The observed heterogeneity and footprint gap may be influenced by factors such as city structure, access, distribution of facilities, and accessibility to different modes of transportation. These findings underscore the importance of mapping and analyzing activity-based carbon footprint in urban sustainability and climate mitigation studies, plans, and actions to enable evaluation of the extent to which urban design and development patterns shape the extent and distribution of activity-based carbon footprint of urban residents.

This study has multiple novel and significant contributions. First, this study is among the first attempts to map and analyze activity-based carbon footprint of urban residents. The majority of the existing literature focuses primarily on consumption-based and home-activity-based carbon footprint dimensions; limited attention has been paid to activity-based carbon footprint and its distribution among urban residents. The findings of this study provide novel and important insights into the extent and distribution of activity-based carbon footprint and the significant gap across different residents. Second, departing from the time-based approach to examining the activity-based carbon footprint of individuals and households from survey data, this study utilizes fine-grained human trajectory data to specify visitations to POIs and distance traveled with high precision and granularity and for a very large sample of urban users. Third, the study and findings show a novel application of fine-grained human mobility and trajectory data for examining urban sustainability problems beyond the current applications in transportation and urban studies. Fourth, the methodology used in this study for data preprocessing and analysis to convert fine-grained human trajectory datasets into POI visitation count and distance traveled provides a computationally efficient approach that could be adopted in future studies. Through these contributions, this study advances the understanding of urban sustainability by better examining human lifestyle and mobility that shape the carbon footprint of urban dwellers. 

The findings from this study also have important implications for decision makers, urban planners, and city managers. By understanding the extent and distribution of users' activity-based carbon footprint resulting from lifestyle patterns and travel behavior, targeted strategies can be developed to reduce the overall carbon footprint. For example, improving access and decentralized facility distribution could reduce both visitation-based and distance-based footprints of urban residents. More equitable distribution of facilities and amenities in low- and medium-income neighborhoods could reduce disparities in carbon footprint and travel distances among users from different income categories. 

This study and its findings also set the stage for future research directions. For example, future studies could examine the extent and distribution of activity-based carbon footprints among users across different cities to specify the extent to which urban form and structure would shape the extent of activity-based carbon footprints and the gaps across different groups of residents. The advancement of understanding of activity-based carbon footprints in cities will move us closer to integrated urban design solutions that would promote sustainability, equity, and climate mitigation.

\section*{Acknowledgments}
This material is based in part upon work supported by the National Science Foundation under CRISP 2.0 Type 2 No. 1832662 grant and the Texas A\&M University X-Grant 699. The authors also would like to acknowledge the data support from INRIX. Any opinions, findings, conclusions, or recommendations expressed in this material are those of the authors and do not necessarily reflect the views of the National Science Foundation, Texas A\&M University, or INRIX.

\section*{Data Availability}
Some of the datasets used in this study are not publicly available under the legal restrictions of the data provider. Interested readers can request waypoint data from INRIX provided here (https://inrix.com/products/) and POI data from SafeGraph from here (https://www.safegraph.com/products/places).

\section*{Code Availability}
The code supporting this study's findings is available from the corresponding author upon request.

\bibliographystyle{unsrt}  
\bibliography{references}  

\end{document}